\documentclass[fleqn,usenatbib]{mnras}
\usepackage{newtxtext,newtxmath}
\usepackage[T1]{fontenc}

\usepackage{xcolor}
\definecolor{c1}{HTML}{02520E}
\definecolor{c2}{HTML}{022C54}
\definecolor{c3}{HTML}{8A0000}

\newcommand{\vect}[1]{\boldsymbol{#1}}

\DeclareRobustCommand{\VAN}[3]{#2}
\let\VANthebibliography\thebibliography
\def\thebibliography{\DeclareRobustCommand{\VAN}[3]{##3}\VANthebibliography}

\usepackage{graphicx}
\usepackage{amsmath}

\title[]{Inferring the density, spin-temperature and neutral-fraction fields of HI from its 21-cm brightness temperature field using machine learning}

\author[B. Bidenko et al.]{
Bohdan Bidenko,$^{1,2}$
L\'eon V. E. Koopmans,$^{1}$\thanks{E-mail: koopmans@astro.rug.nl}
P. Daniel Meerburg$^{2}$\vspace{0.1cm}
\\
$^{1}$Kapteyn Astronomical Institute, University of Groningen, PO Box 800, NL-9700 AV Groningen, the Netherlands\\
$^{2}$Van Swinderen Institute for Particle Physics and Gravity, University of Groningen,
Nijenborgh 4, 9747 AG Groningen, The Netherlands}

\date{Accepted XXX. Received YYY; in original form ZZZ}

\pubyear{XXXX}

\begin{document}
\label{firstpage}
\pagerange{\pageref{firstpage}--\pageref{lastpage}}
\maketitle

\begin{abstract}
The 21-cm brightness-temperature field of neutral hydrogen during the Epoch of Reionization and Cosmic Dawn is a rich source of cosmological and astrophysical information, primarily due to its significant non-Gaussian features. However, the complex, nonlinear nature of the underlying physical processes makes analytical modelling of this signal challenging. Consequently, studies often resort to semi-numerical simulations. Traditional analysis methods, which rely on a limited set of summary statistics, may not adequately capture the non-Gaussian content of the data, as the most informative statistics are not predetermined.
This paper explores the application of machine learning (ML) to surpass the limitations of summary statistics by leveraging the inherent non-Gaussian characteristics of the 21-cm signal. We demonstrate that a well-trained neural network can independently reconstruct the hydrogen density, spin-temperature, and neutral-fraction fields with cross-coherence values exceeding 0.95 for $k$-modes below $0.5$ Mpc h$^{-1}$, based on a representative simulation at a redshift of $z \approx 15$. To achieve this, the neural network utilises the non-Gaussian information in brightness temperature images over many scales. We discuss how these reconstructed fields, which vary in their sensitivity to model parameters, can be employed for parameter inference, offering more direct insights into underlying cosmological and astrophysical processes only using limited summary statistics of the brightness temperature field, such as its power spectrum.
\end{abstract}


\section{Introduction \label{sec:intro}}

Throughout much of the history of the evolution of the universe in its first billion years, its baryonic component consisted mostly of neutral hydrogen. Analysis of physical processes during this period is possible with the help of the 21-cm emission line caused by the spin-flip transition of the electron-pair making up the hydrogen atom. 

The unique physical information contained in this signal has led to the development of an entire field known as 21-cm cosmology \citep[e.g.,][]{2001PhR...349..125B,2006PhR...433..181F,2012RPPh...75h6901P, Mesinger2019}. Forecasts for experiments such as the GMRT \citep{Paciga2013}, LOFAR \citep{Patil2017, mertens20, 2020MNRAS.493.4728G}, NenuFAR \citep{Mertens2021, Munshi2023}, LWA \citep{Eastwood_2019}, MWA \citep{barry19, li19, Trott2020}, PAPER \citep{kolopanis19}, HERA \citep{DeBoer_2017, Abdurashidova_2022}, and SKA \citep{2015aska.confE...1K, 2015aska.confE..10M} aim to significantly improve constraints on the models describing the astrophysics of the first ionizing sources as well as cosmological parameters \citep{2020PASP..132f2001L}.

The traditional approach to studying the 21-cm signal has relied on measuring or constraining fluctuations of the 21-cm brightness temperature, which can be qualified by taking moments of the distribution, of which the most well-studied example is the power spectrum \citep[see e.g.,][]{Morales2004}. Current attempts to measure the 21-cm power spectrum (see references above) mostly aim to detect a signal from the period when the universe was completely neutral to mostly ionized. This window is referred to as Cosmic Dawn (CD) and the Epoch of Reionization (EoR) and roughly spans $30\leq z \leq 5$. Below $z = 5$, the 21-cm signal can still be observed inside of halos/galaxies as a biased tracer of structure on large scales, which is the target for current experiments such as CHIME \citep{CHIME:2022dwe}, Tianlai \citep{2012IJMPS..12..256C}, HIRAX \citep{Crichton:2021hlc} and MEERKAT \citep{Cunnington2022}, who recently claimed detection of the 21-cm signal power spectrum at low redshift. 

While most effort has been put into detecting the power spectrum (or cross-power with large scale structure in the case of low-$z$ 21-cm signal, see e.g. \citep{Pen:2008fw, Chang:2010jp, Masui:2012zc, Anderson:2017ert, eBOSS:2021ebm}), the distribution of 21-cm fluctuations is non-Gaussian and its power spectrum does not capture {\sl all} information present in the field. As such, relying on the power spectrum alone is sub-optimal in constraining the physics that sources the signal. In this paper, we focus on the late stages of the CD or early stages of the EoR, when the universe is still mostly neutral, but patches of ionized regions start to appear. Several studies have explored the non-Gaussian nature of the signal in this epoch, both using higher-order moments, such as the bi-spectrum \citep{Saxena:2020san, Hutter:2019yta, Watkinson:2020zqg, Watkinson:2021ctc}, as well as the morphology \citep{Chen2019, Gazagnes:2020iwr, Giri2021, Kapathia2021}. The bi-spectrum is the most sensitive when non-Gaussianity is weak, which is not the case during the CD and EoR. 

What complicates the application of summary statistics is the lack of a closed analytical formulation of the fluctuations during this epoch (see, for example, \citet{McQuinn:2018zwa, 2022PhRvD.106l3506Q} for recent attempts). This stands in contrast to the post-EoR epoch and the 21-cm signal from the Dark Ages. Thus, both the modelling and analysis of the signals during the CD and EoR epochs heavily rely on simulations. Consequently, the non-linear nature of the processes governing astrophysics and cosmological evolution, as well as the resulting non-Gaussian signal, are challenging to qualify analytically. Developing a standardised estimator that optimally captures this information is likely impossible. Fortunately, with the advancement of machine learning (ML), it has become possible to infer complex relationships between different datasets, enabling us to qualify data beyond its summary statistics.

Applying ML techniques to cosmological and astrophysical problems is a rapidly developing field. For example, in the context of 21-cm data, deep-learning-based methodologies were shown to be successful in identifying ionised regions in the presence of challenging astrophysical foreground and instrumental noise contamination \citep{2021MNRAS.504.4716G, 2021MNRAS.505.3982B, 2024MNRAS.528.5212B}, in applying sample generation and parameter inference \citep{Schmit_2017, Zhao_2022,2020MNRAS.493.5913L}  and reconstructing the density fields  \citep{2021ApJ...907...44V}. 

One can think of the 21-cm field during the CD and EoR as a product of cosmological and astrophysical evolution \citep[][]{Madau2004}. The key ingredients of the 21-cm brightness signal are the spin temperature $T_S$, the ionization fraction $x_{\rm HI}$ and baryonic matter density fluctuations $\delta_b$. Optimal use of the data for astrophysical and cosmological inference would benefit from separating astrophysical quantities $T_S$ and $x_{\rm HI}$ from the cosmological signal $\delta_b$, and vice versa. The same reasoning applies when considering the astrophysics hidden in the signal. It is known that at the level of the power spectrum, parameters that are associated with cosmology and astrophysics can have strong degeneracies, which can only be partially broken when considering measurements of the power spectrum at several redshifts \citep[see e.g.,][]{10.1093/mnras/stx2118}. Given the non-Gaussian nature of the 21-cm signal during the CD and EoR, it is only natural to explore whether non-Gaussian information is capable of reducing these degeneracies and providing additional information. This is the main motivation behind this paper. 

In particular, inspired by applications of machine learning aimed at density-field reconstruction \citep[][]{2024JCAP...02..031F}, we develop a three-dimensional field-to-field reconstruction network that is capable of reconstructing the three fields $T_S$, $x_{\rm HI}$ and $\delta_b$ directly from the brightness fluctuations $\delta T_b$, i.e. the observed field. The network utilizes the full non-Gaussian information in the observed field to reconstruct the other fields. We show that the information in the power spectra of these fields is complementary to the power spectrum of $\delta T_b$, and if combined, they could lead to improved parameter constraints. In this paper, we consider a simple set-up focused on a single redshift, but considering a range of reionization histories simulated using the semi-numerical public code \verb|21cmFAST| \citep{10.1111/j.1365-2966.2010.17731.x, Greig2015, Park_2019}. Performing parameter forecasts and exploring applications of the reconstructed fields will be presented in future publications. As such, this paper serves as a proof-of-principle.    

Our paper is organized as follows. 
In section \ref{sec:si}, we briefly describe the difference between the observable 21-cm signal and the underlying physical fields and motivate using a simple two-parameter Fisher forecast why reconstructing the field could benefit parameter inference.
In section \ref{sec:me}, we describe the simulations used for training and testing of the neural network and provide details on its architecture and the training procedure. In section \ref{sec:re}, we present the results of the reconstruction and qualify the performance of the network by measuring the cross-spectra. Finally, we discuss limitations and discuss future directions in section \ref{sec:co}.

\section{The 21-cm signal of Neutral Hydrogen} \label{sec:si}

In this section, we summarise the brightness temperature field's dependence on the neutral hydrogen fraction, its spin temperature, and the baryonic overdensity field. Through a straightforward Fisher information analysis, we demonstrate that the power spectra of these fields provide additional information beyond that contained in the power spectrum of the brightness temperature field alone. 

\subsection{The Brightness Temperature Field}

The observed 21-cm brightness temperature is defined by the difference between the spin temperature $T_S$ and the Cosmic Microwave Background (CMB) temperature $T_{\rm CMB}$, modulated by the Sobolov optical depth along the line of sight. 
It can be approximated as follows \citep{Madau2004, 2006PhR...433..181F}:

\begin{eqnarray}
        \delta T_b&=&\frac{T_S-T_{\rm CMB}}{1+z}(1-{\rm e}^{-\tau_\nu})\nonumber\\
&\approx&27  x_{\rm{HI}}\left(1+\delta_b\right)\left(\frac{\Omega_bh^2}{0.023}\right)\left(\frac{0.15}{\Omega_mh^2}\frac{1+z}{10}\right)^{1/2} \nonumber\\
&&\times\left(\frac{T_S-T_{\rm CMB}}{T_S}\right)\left[\frac{\partial_r v_r}{(1+z)H(z)}\right]\,\rm{mK}
    \label{eq:tb},
\end{eqnarray}

Here, \(x_{\rm{HI}}\) is the neutral-hydrogen fraction, \(\delta_b\) is the fractional overdensity of baryonic matter, and \(\partial_r v_r\) is the velocity gradient along the line of sight. The spin temperature measures the number of hydrogen atoms in the two hyperfine spin-states (parallel or anti-parallel). During the initial period after the CMB has formed, the CMB and the gas temperature are tightly coupled, and a balance of collisional and radiative transitions is set, \(T_S = T_{\rm CMB} = T_g\). As the universe expands, the CMB photons decouple from the gas while collisions couple the spin temperature to the gas temperature, which cools adiabatically. This is the moment (\(z \sim 200\)) when the 21-cm brightness can be seen in absorption against the CMB. As the gas dilutes further, collisions become inefficient, and the spin temperature approaches the CMB temperature (\(z \sim 50\)). When the first stars form, Lyman-$\alpha$ photons again couple the spin temperature to the gas, leading to a peak in the brightness temperature (still in absorption) at \(z \sim 20\), until X-rays from the first stars heat the gas during the EoR. The brightness temperature can then be observed in either absorption (\(T_S < T_{\rm CMB}\)) or emission (\(T_S > T_{\rm CMB}\)). During most of the EoR, \(T_S > T_{\rm CMB}\), and hence the brightness temperature is positive. This is the regime we will study in this paper. We refer to \cite{2012RPPh...75h6901P} and \cite{Mesinger2019} for a more detailed review of the above processes.

The primary focus of nearly all studies has been the summary statistics of the $\delta T_b$ field. When observing the 21-cm signal, the $\delta T_b$ field is the only direct observable. As shown by Eq.~\eqref{eq:tb}, however, the brightness temperature is related to three other fields $T_S$, $x_{\rm HI}$ and $\delta_b$. Note that in principle, it also couples to fluctuations in the CMB temperature, but these are typically negligible since $\Delta T_{\rm CMB}/T_{\rm CMB} \sim \mathcal{O}(10^{-5})$. The evolution of the fluctuations in these fields is determined by complex cosmological and astrophysical processes. The baryonic matter density perturbations, $\delta_b$, grow through gravitational interactions but, on small scales, are expected to be affected by astrophysical feedback effects. The neutral fraction of hydrogen  (and its fluctuations), $x_{\rm{HI}}$, is determined by photoionization by early stars and quasars. The spin temperature field, $T_S$, is defined by collisional coupling in dense regions, coupling with CMB radiation in lower kinetic temperature regions, X-ray heating, and Ly-$\alpha$ pumping mechanism \citep[for details see e.g.][]{2001PhR...349..125B,2006PhR...433..181F,2012RPPh...75h6901P, Mesinger2019}. 

\begin{figure}
 \includegraphics[width=\columnwidth]{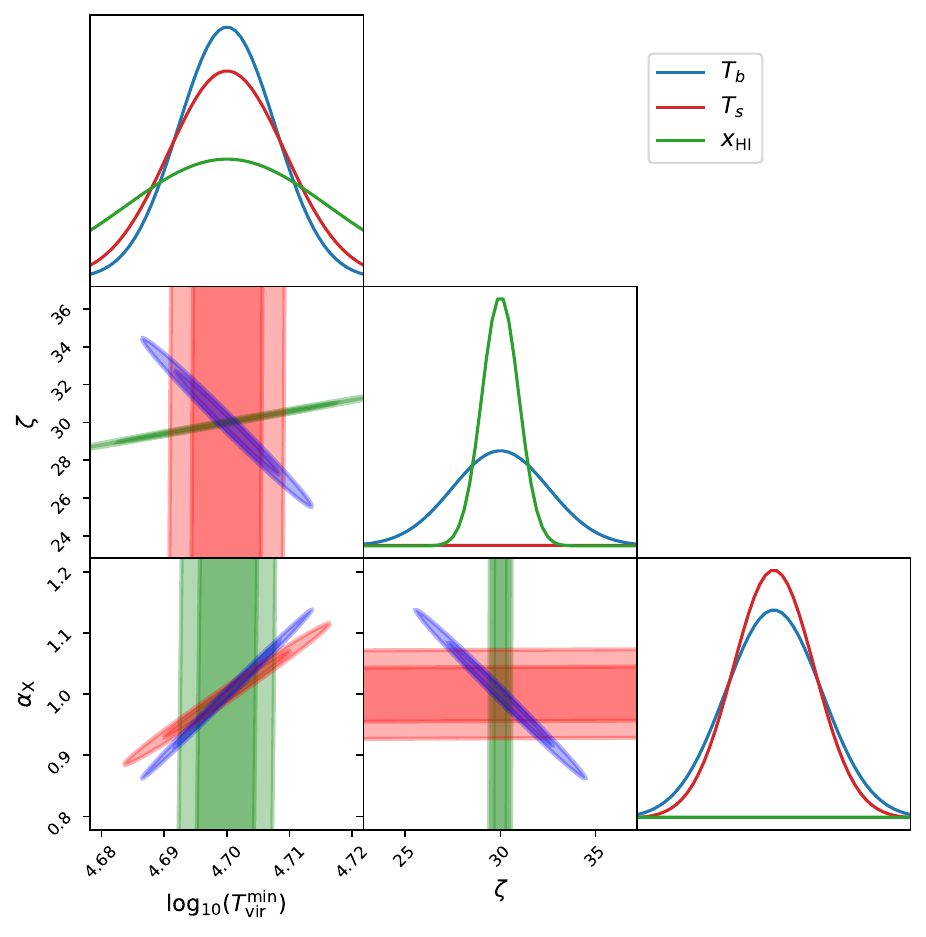}
 \caption{Fisher forecast for $\zeta$, $\alpha_{\rm X}$ and $T_{\rm vir}$ derived from the power spectrum of the brightness temperature field (blue), the spin temperature field (red), and the ionization field (green). We used cosmic variance limited power spectra in the range $0.04\rm{~h/Mpc} \leq k\leq 0.6 \rm{~h/Mpc}$. The power spectra were obtained from simulations at $z = 15$ obtained with the semi-numerical code \texttt{21cmFAST}, fixing all other parameters. The derivatives were computed using finite difference derivatives. This simple Fisher analysis shows how different fields respond to reionization parameters and whether the information in these fields is independent. Combining their information would break degeneracies and improve parameter constraints.}
 \label{fig:fisher}
\end{figure}

Modelling these effects analytically is challenging \citep[see e.g.][]{McQuinn:2018zwa, 2022PhRvD.106l3506Q} and we rely on numerical simulations to make predictions about the brightness temperature field, which relies on $T_S$, $\delta_b$ and $x_{\rm HI}$. Full numerical codes that include ray-tracing are computationally expensive, and to render the number of cosmological boxes necessary for parameter inference, or in this case, for the training of a neural network, is infeasible at the moment. Hence, for that purpose, we use the semi-numerical code \verb|21cmFAST| \citep{10.1111/j.1365-2966.2010.17731.x, Park_2019}. It starts with cosmological initial conditions rendered using 2LPT using standard cosmological parameters. The code models the formation and evolution of dark-matter halos using the Press-Schechter formalism, incorporates star formation and feedback processes within these halos, and calculates the propagation of ionizing radiation from stars and galaxies. Key parameters are the ionizing efficiency $\zeta$, representing the number of ionizing photons emitted per baryon in collapsed structures and $T_{\rm vir}$,  which sets the conditions under which gas can collapse and form stars within dark matter halos and $\alpha_{\rm X}$ which quantifies how efficiently X-ray photons can heat the surrounding intergalactic medium (IGM) compared to ionizing photons produced by stars and galaxies.

\subsection{Fisher Information}

In this subsection, we show the information these fields contain on three astrophysical parameters to motivate our effort to reconstruct the fields determining the observed brightness temperature.  

In Figure~\ref{fig:fisher}, we show the one- and two-dimensional marginal constraints determined by performing a simple Fisher forecast \citep{dafebb01-f6c3-3bb0-8577-2b844df40cce} derived directly from \verb|21cmFAST| simulations using the power spectra of $\delta T_b$ (blue contours), $T_S$ (red) and $x_{\rm{HI}}$ (green). Because we focus on astrophysical parameters rather than containing cosmology \citep[as done by][]{2024JCAP...02..031F}, we can ignore the baryonic density field. The results are obtained for a noiseless cosmic-variance-limited experiment, and the power spectra are measured at redshift $z=15$ using 256$^3$ resolution boxes with a side scale of 1024 Mpc/h, ensuring $T_{\rm CMB}/T_S$ is not so small such that the effect of $T_S$ is negligible.

We notice that the X-ray spectral index parameter, $\alpha_{\rm X}$, has stronger constraints from the $T_S$ power spectrum than from $\delta T_b$ due to stronger $T_S$ coupling with the X-ray background \citep{1959ApJ...129..536F,10.1111/j.1365-2966.2010.17731.x}. 
The parameter that defines the ionizing efficiency of  UV radiation from early galaxies $\zeta$ is constrained stronger by the $x_{\rm HI}$ field.
The minimal virial temperature parameter $\log_{10} (T^{\rm min}_{\rm vir})$ has similar constraints from $T_S$ and $\delta T_b$. However,
due to different types of degeneracies with the $\alpha_{\rm X}$ parameter, the constraints could be optimised by combining information from these fields (after properly accounting for their covariance). This simple parameter Fisher forecast shows that if one were to have access to these additional fields, a simple summary statistic, such as their power spectra, could already improve astrophysical parameter constraints.

However, since we do not have direct access to these fields, we explore 
the potential of a deep learning-based algorithm for the reconstruction of the $T_S$, $x_{\rm{HI}}$ and $\delta_b$ using only the observed $\delta T_b$ field. Note that the information to reconstruct these fields must be derived from the non-Gaussian nature of the brightness temperature. In principle, all information is present in this one observable field, but this information is contained in multiple higher-order moments. If we were able to capture and model that information correctly, the resulting constraints would be more optimal. However, given the challenge of both modelling and measuring these higher-order moments, a direct reconstruction would allow for a relatively straightforward analysis using the power spectra {\sl after} reconstruction. 

\begin{figure*}
 \includegraphics[]{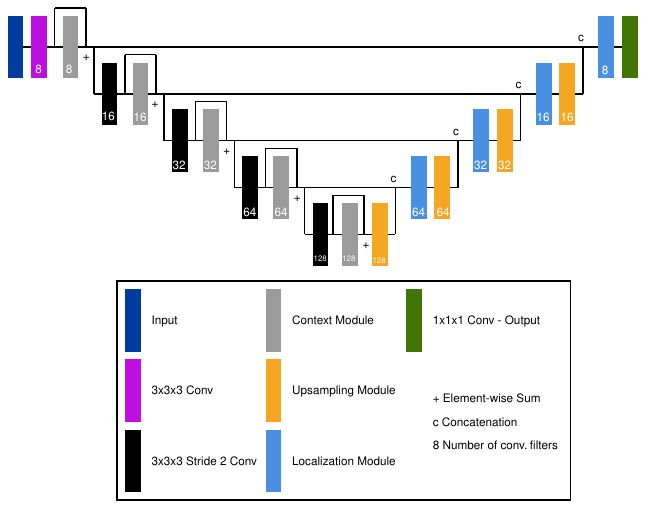}
 \caption{Architecture of the neural network used for reconstruction. Image credit: \citet{2024JCAP...02..031F}}
 \label{fig:unet}
\end{figure*}

\section{Methodology}

In this section, we shortly describe the simulations of the brightness-temperature field and the neural network used to infer the neutral-fraction, baryonic-density and spin-temperature fields from the brightness-temperature field.

\label{sec:me}
\subsection{Simulations}
\label{sec:me:1}

We aim to recover underlying physical fields using supervised machine learning based on the information present in the observable. As a first attempt, we will use simulations without including any observational and instrumental effects. Therefore, the simulations used in this work do not contain any noise, and the information about the physical processes is limited only by the quality of the simulations, their size, and the binning effect. We leave that to a future analysis. We simulate the four fields using {\tt 21cmFAST}.
For the training and validation set, we generate simulations with different astrophysical parameters: $\zeta \in \{28, 32\}$, $\alpha_{\rm{x}} \in \{0.935 ,1.065\}$, and $\log_{10}(T_{\rm{vir}}^{\rm{min}}) \in \{4.67, 4.69, 4.71, 4.73 \}$.  
We chose parameters whose constraints could benefit from the field reconstruction as shown in Figure \ref{fig:fisher}. 
The chosen range of parameters lies within $1\sigma$ of the parameter constraints forecast for upcoming experiments \citep[e.g.][]{2023MNRAS.525.6097S}.
The test data is generated with parameter values between the test grid points: $\zeta = 30$, $\alpha_{\rm{x}} = 1$, and $\log_{10}(T_{\rm{vir}}^{\rm{min}}) = 4.7$, which corresponds to the ``FAINT GALAXIES'' model of \cite{10.1093/mnras/stx2118}. With this set of simulations, we determine to what extent our network can interpolate to models that are not part of the training set. A more elaborate study of reconstruction quality as a function of parameter space is a topic of ongoing investigation. The range considered here should again serve as a proof-of-principle. 

We chose to study the reconstruction in a regime close to the maximum average brightness temperature. Therefore, we simulate all fields at a redshift $z=15$. Each simulation data cube has a resolution of $256^3$ voxels and a physical side size of 1024 Mpc/h. 
Besides $\alpha_{\rm X}$, $T_{\rm vir}$ and $\zeta$, all other astrophysical and cosmological parameters in our simulations were set to the default values of {\tt 21cmFAST}. All simulations used for training, validation and testing are generated with different randomised initial realisations (seeds).

\begin{figure*}
 \includegraphics[width=1\textwidth]{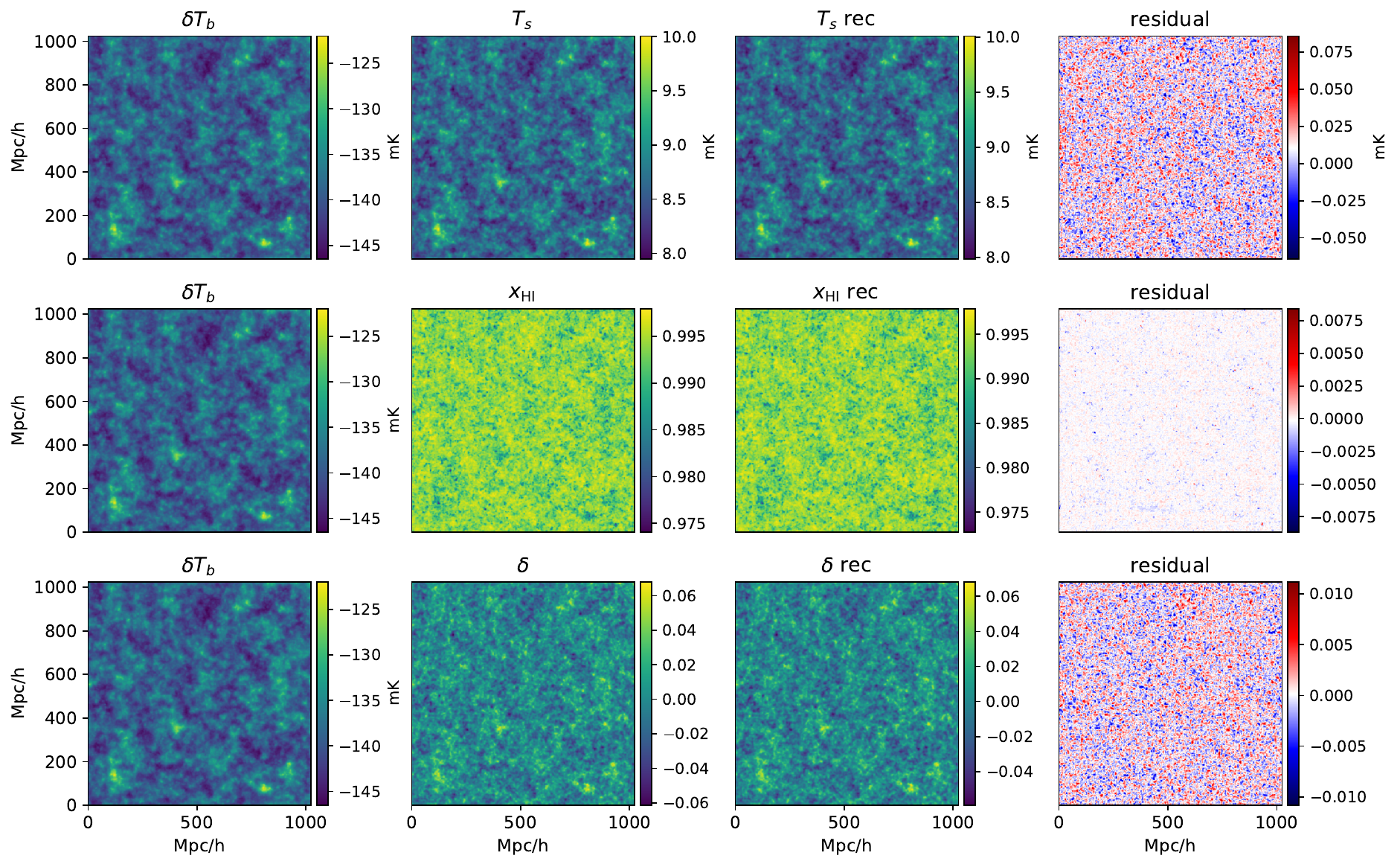}
 \caption{Slices of the four fields projected along the third dimension of the data cube.
 First column: the input brightness temperature field. Second column: corresponding target field. Third column: reconstructed field. Fourth column: the difference between the reconstructed and true field. 
 The first row shows the reconstruction of the spin temperature field $T_S$, the second row shows the reconstruction of the ionization fraction $x_{\rm HI}$, and the third row shows the reconstruction of the baryonic matter fractional overdensity $\delta_b$. Residuals are very small, indicating high-quality reconstruction for all three target fields.}
 \label{fig:res}
\end{figure*}

\subsection{U-net architecture \label{sec:intro:rec}}

For the reconstruction of the physical fields, we use the neural network with the same architecture as was proposed in \citet{2024JCAP...02..031F}, which was inspired by the work of 
\citet{2021MNRAS.504.4716G}. The reconstruction algorithm starts with the construction of six additional fields\footnote{Although they are three-dimensional cubes, we conventionally call them {\sl fields} throughout this work.} for each input $\delta T_b$ field:
\begin{eqnarray}
\vect{u}(\vect{x}) = \int d^{3} \vect{k} \frac{-i \vect{k}}{k^2} \delta T_b (\vect{k}) e^{i\vect{k}\cdot\vect{x}} \label{eq:ps1} \\
\partial_i(\vect{u})_{j}(\vect{x}) = \int d^{3} \vect{k} \frac{k_i k_j}{k^2} \delta T_b(\vect{k}) e^{i\vect{k}\cdot\vect{x}}, \label{eq:ps2}
\end{eqnarray}
where $\delta T_b(\vect{k})$ is a Fourier transform of the temperature brightness field.
It was shown by \citet{2024JCAP...02..031F} that adding the gradient fields improve reconstruction. 

The neural network architecture utilised in our study consists of an initial convolutional layer with a $3\times3\times3$ kernel size, followed by a context block consisting of two additional convolutional layers with the same kernel size. A residual connection is then applied, combining the output of the initial convolution with that of the context block. This process is repeated, with the initial convolutional layer now employing a stride of two, resulting in the down-sampling of feature maps by a factor of two in each spatial dimension. This down-sampling procedure is iterated three more times until the feature maps reach a size of $16\times16\times16$ cells. Subsequently, the feature maps are upsampled by a factor of two, and the outputs of the residual connections from the down-sampling phase are concatenated, creating skip connections. These concatenated feature maps are then fed into a localisation block, comprising a $3\times3\times3$ ordinary convolutional layer followed by a $1\times1\times1$ convolutional layer, with this process repeated until the input size reaches $256\times256\times256$. Finally, one more convolutional layer of size $1\times1\times1$ is applied to generate a single feature map representing the reconstructed field. Throughout the architecture, except for the final convolutional layer, each layer consists of convolution, instance normalisation, and leaky ReLU activation functions.

\subsection{Training}

The U-net is trained independently for each type of the reconstructed physical field due to memory limitations. Thus, we train three independent neural networks in this work.
For each network, we use $240$ simulations of $\delta T_b$ field data cubes as input dataset and corresponding $\delta_b$, $T_s$, and $x_{\rm{HI}}$ as output datasets. The training set contains $15$ field realisations for each point in parameter space.  
These simulations are used to optimise neural network parameters using a mean square error loss function:
\begin{eqnarray}
    {\cal L}_{\rm{MSE}} = \frac{1}{N_{\rm{sim}}}\frac{1}{256^{3}} \sum_{i=1}^{N_{\rm{sim}}}\left(F_{i,\rm{true}}-F_{i,\rm{rec}}\right)^{2},
    \label{eq:loss}  
\end{eqnarray}
where $F_{i,\rm{true}}$ and $F_{i,\rm{rec}}$ are the true target and reconstructed field, respectively, and the difference is calculated between individual voxel values. The loss function is normalised by the number of voxels and simulations in the set $N_{\rm sim}$. 
The validation dataset consists of $48$ simulations (three per parameter-space point), and the test dataset consists of $90$ simulations with one fixed parameter set as discussed in section \ref{sec:me:1}.

For the spin temperature field, $T_S$, we cap the field's maximum value at the 99.5th percentile. This limit is essential to prevent instability in the learning process that arises when the U-net reconstructs very large localised peaks and affects only a few voxels. This adjustment has a negligible impact on the power spectrum and other statistical properties of the fields. Finally, for training purposes, we normalize fields using the standard deviation of the training set fields and subtract the mean value of the training fields. The true values of the reconstructed fields are recovered through inverse re-scaling and applying a corresponding shift. 

\section{Results}

In this section, we present a representative reconstruction of the neutral-fraction, baryonic-density, and spin-temperature fields from a random brightness-temperature field, which was given as input to the trained U-net. We visually inspect these, analyse their power spectra, and assess their cross-coherence with the true underlying fields. Again, we emphasise that this is proof of concept, and a full exploration of a wider range of redshifts and model parameters is left for future analyses.

\label{sec:re}
\subsection{Visual inspection}

To qualitatively assess the reconstructed fields, we first perform a visual comparison. Representative examples of the network reconstruction are shown in Figure~\ref{fig:res}. In the first column, we show a slide of the input $\delta T_b$ field along one of the data cube's axes. We note that these are the same for each row but shown for ease of comparison with the other fields. The second column shows the true target fields $T_S$, $x_{\rm HI}$, and $\delta_b$ in the first, second and third rows, respectively. The third column shows the results of the reconstruction, obtained with a U-net for each field. The difference between the true and the reconstructed field is shown in the fourth column.

Remarkably, there are no strong visual imperfections in residuals across all the simulations used in training, validation and tests. Residuals do not feature apparent structure or inhomogeneities. The $x_{\rm HI}$ residuals show some minor localised peaks. These peaks are associated with a small misalignment of the positions of ionised regions between the reconstructed and the true field. However, due to the small-scale nature of these features, they do not introduce significant errors on the scales that forthcoming experiments are expected to probe with reasonable signal-to-noise (i.e., $k\leq0.5$ h/Mpc).

\subsection{Power spectra of the reconstructed and true fields}

\begin{figure*}
 \includegraphics[width=1\textwidth]{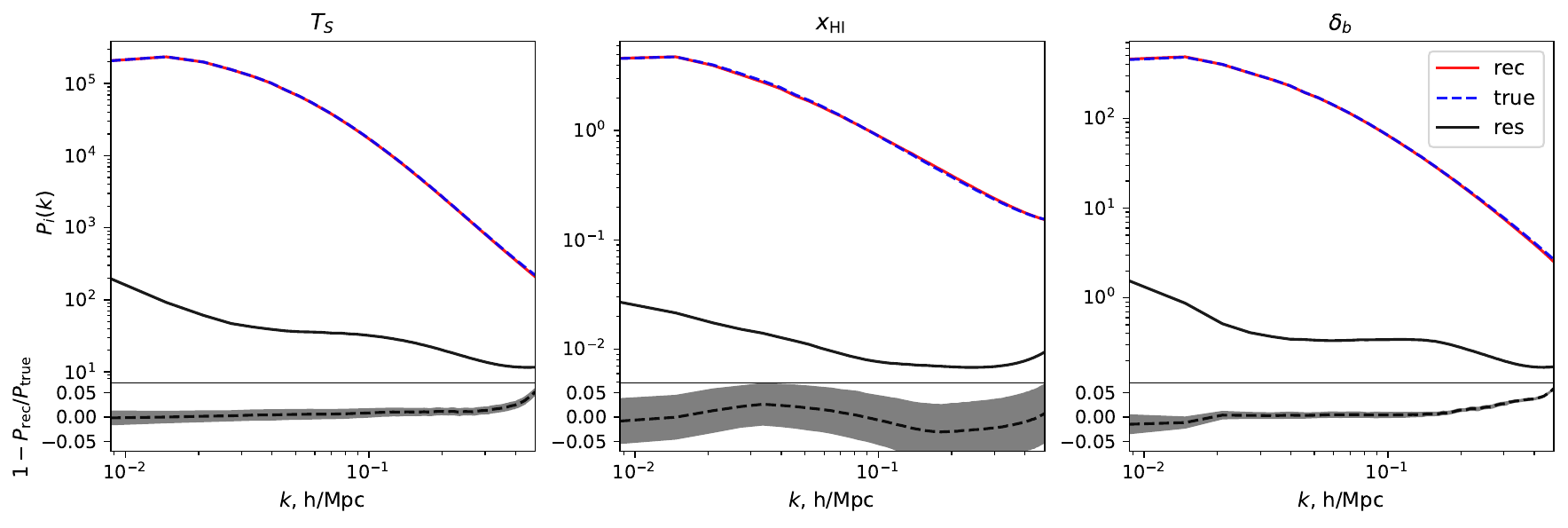}
 \caption{Top panel: Power spectrum of reconstructed field (red), true target field (blue dashed), and their residual (black). Left: spin temperature field; central: neutral hydrogen fraction; right: baryonic density fluctuations. Bottom panels: fractional deviation of the reconstructed field power spectrum from the target true field power spectrum.}
 \label{fig:pk}
\end{figure*}

To quantitatively assess the reconstructed fields, we compute the power spectrum of the reconstructed field and compare it to the power spectrum of the target field. The power spectrum of the binned three-dimensional fields is calculated using the following estimator:
\begin{eqnarray}
    {P}(k_i) =  \langle |F(\vect{k})|^2 \rangle = \frac{1}{N_{k_i}} 
    \sum_{\vect{k} \in k_{i}\pm \Delta k} |F(\vect{k})|^2,
    \label{eq:pk}
\end{eqnarray}
where $F(\vect{k})$ is a Fourier transform of the field $F$. The square magnitude of the Fourier transform is calculated as the spherical layer average in the wavenumber space defined by the radius $k_i$, with a width $\Delta k$ equal to the fundamental Fourier mode of the data cube. The number of Fourier modes in the spherical layer $N_{k_i}$ is taken into account as a normalization factor.

We present the power spectra in Figure~\ref{fig:pk}, averaged over the test dataset. From left to right, the panel shows the statistical properties of the $T_S$ field, the $x_{\rm HI}$ field, and $\delta_b$ field, respectively.
The power spectra of the reconstructed fields (shown in red lines) in all three cases deviate from true target power spectra (blue dashed lines) by less than $10\%$ across the entire range of scales.
We also show the residual power spectra (black lines). 
The residual power is $\leq7\%$ of the field power spectrum for $k\leq 0.5~\rm{h/Mpc}$.
The $x_{\rm HI}$ residual power spectrum increases on small scales (high $k$ values) as expected from the localized peaks in the residual.
Despite this behaviour of the residual, the deviation of the reconstructed power spectrum is $\leq 2.5\%$ in the case of $x_{\rm HI}$ field (black dashed line in the bottom panel of Figure~\ref{fig:pk}).
Similarly, the reconstructed power spectra of $T_S$ and $\delta_b$ deviate $\leq 7\%$. Overall, the reconstructions are remarkably good.

\subsection{Cross-coherence between the reconstructed and true fields}

While the previous section showed that the U-net reconstruction method successfully recovers the power spectra of the target fields, a more ambitious goal is to recover the target fields on a pixel-to-pixel level. This is possible because the network relies on a pixel-by-pixel comparison of the loss function used in training (Eq.~\eqref{eq:loss}).
To further quantify the quality of the reconstructed fields, we compute the cross-coherence between the target (i.e. true) and the reconstructed field. Similar to the power spectra, the coherence is defined as the normalised cross-power spectrum, as follows:
\begin{eqnarray}
    C_{i,j}(k) = \frac{\langle F_{i}(\vect{k})F^*_{j}(\vect{k}) \rangle^{2}}{\langle |F_{i}(\vect{k})|^2 \rangle \langle |F_{j}(\vect{k})|^2\rangle}=\frac{P_{i,j}^{2} (k)}{P_{i}(k)P_{j}(k)}.
\end{eqnarray}
Here $P_{i,j}$ is cross-power spectrum of field $i$ and field $j$, $P_{i}$ and $P_{j}$ are auto-power-spectra of fields $i$ and $j$, obtained using Eq.~\eqref{eq:pk}, respectively. Coherence values of unity correspond to identical target and reconstructed fields, and values of zero correspond to completely uncorrelated target and reconstructed fields.

\begin{figure*}
 \includegraphics[width=1\textwidth]{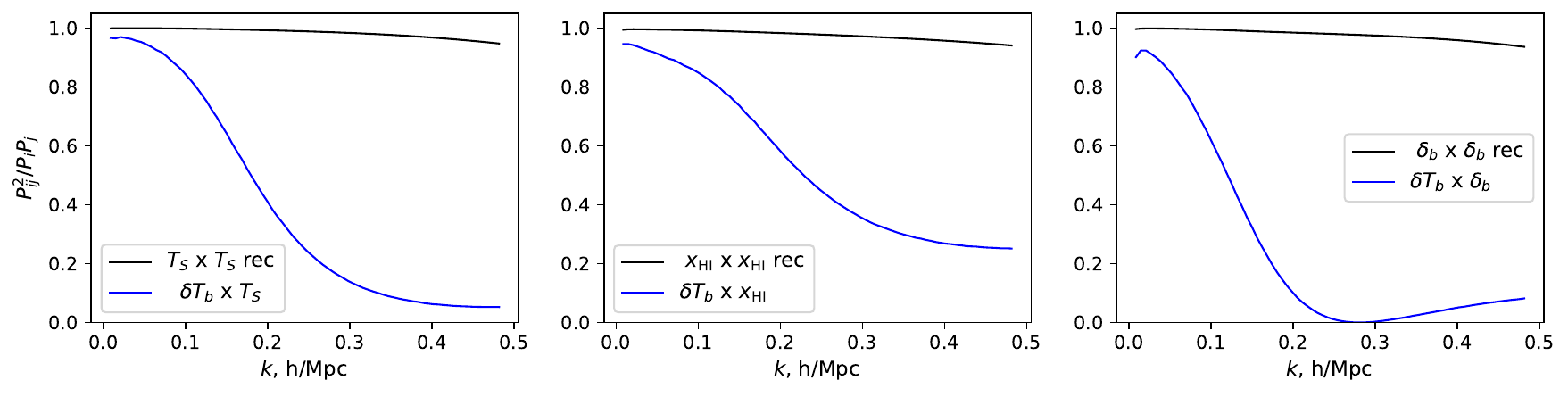}
 \caption{Cross-correlation (coherence) between the target field from the test dataset brightness temperature fields (blue) and the reconstructed field (black). The left panel is the spin temperature field; the central panel is the hydrogen neutral fraction; the right panel is the density field fluctuations. Curves represent the mean value for the entire test dataset. On large scales, the brightness temperature is coherent with all other fields. On small scales, the brightness temperature becomes more decoherent. After reconstructing the target fields, we obtain significant coherence up to $k \sim 0.5$ Mpc/h.}
 \label{fig:xcorr_all2}
\end{figure*}

We show the coherence between reconstructed and target fields (black lines), as well as the coherence of the target with the input field $T_b$ (blue), in Fig.~\ref{fig:xcorr_all2}. The reconstructed fields in the test dataset have a mean coherence of $\geq 0.92$ for scales $k\leq 0.5 \mathrm{h/Mpc}$ with the target field. The coherence increases on larger physical scales, which is in agreement with the trend in the power spectrum residual. It is also indicative that non-Gaussian information is being used to reconstruct these scales. The rapid decrease in cross-coherence on smaller scales ($k\geq 0.15 \mathrm{h/Mpc}$) between the reconstructed field and target field shows that they are not simply scaled versions of each other. 

To test that a single point in the reconstructed field depends on a wide range of scales and surrounding points, we examined the network's reconstruction capabilities by applying cuts in Fourier space, thereby removing certain scales from the input field similar to \cite{2021MNRAS.504.4716G}. As anticipated, the quality of the reconstruction deteriorates, showing sensitivity to the specific scales being excluded. When large scales (small $k$ values) are removed, the networks can still reconstruct modes absent in the observed fields, indicating the importance of small, non-Gaussian scales in the reconstruction process. Conversely, removing small scales hampers the reconstruction, as it eliminates critical non-Gaussian information. A comprehensive analysis of the specific scales contributing to the reconstruction will be addressed in future work.

\section{Discussion and Conclusions}
\label{sec:co}
We have presented a deep learning-based (U-net) method that successfully reconstructs all three physical fields (i.e., baryonic density, neutral fraction and spin-temperature) that largely determine the observed 21-cm brightness temperature field of neutral hydrogen during the Epoch of Reionization and Cosmic Dawn. The motivation to build such a network stems from the fact that these fields have different responses to both astrophysical and cosmological parameters, making it possible to independently harvest the information in these fields to improve parameter constraints. The network utilises the non-Gaussian information present in the brightness temperature field. This non-Gaussian information non-trivially couples modes at different scales and the networks use this coupling to reconstruct the target fields. This is most evident from the fact that the observed field $T_b$ has poor cross-correlation with the target fields on small scales, while after reconstruction, we find very high cross-correlation on those same scales. Since this is not simply an overall re-scaling of the field, the information is recovered from non-linear relations, i.e. from non-Gaussian information present in the brightness temperature data cube. Concretely, we find the following:
\begin{itemize}
\item The power spectra of the reconstructed spin temperature $T_S$,  neutral hydrogen fraction $x_{\rm HI}$ and baryonic matter density perturbation $\delta_b$ field -- simulated with the semi-numerical code \verb|21cmFAST| -- deviate $\leq 7\%$ at $k=0.5$ h/Mpc from the power spectra of the true target field, and even less on larger scales. 

\item The cross-coherence of the reconstructed fields with the true target fields is found to be $C\geq0.92$ on scales $k<0.5 \rm h/Mpc$, much higher than the cross-coherence with the brightness-temperature field. This confirms that the reconstruction extends beyond the similarity of statistical properties but features a high degree of pixel-to-pixel accuracy of the reconstructed field.

\end{itemize}

The primary conclusion of our proof-of-principle analysis is that the redshifted 21-cm brightness temperature field retains most of the information regarding the three underlying fields: baryonic matter density, neutral fraction, and spin temperature. These fields can be accurately reconstructed using a neural network that effectively disentangles them from a single input field. 

However, several additional steps are necessary to fully evaluate the practical applications of these methods. First, it is crucial to assess whether the inclusion of realistic noise, instrumental effects, and foregrounds impacts the reconstruction quality. Second, performing a Fisher analysis on the reconstructed fields is essential to determine their utility for parameter inference, as demonstrated by \citet{2024JCAP...02..031F}. Third, our current setup focuses on a relatively high redshift of $z=15$, where the ionized regions are small\footnote{While there is still neutral hydrogen in these regions, it will be undetectable with current and forthcoming 21-cm experiments}. Some preliminary tests, however, show that the reconstruction quality diminishes when large ionized patches are present in the observed field. To mitigate this, one possible approach is to complement the 21-cm brightness measurements with other line tracers, such as {\tt OIII} or {\tt CO}, which exhibit significant fluxes in ionised regions. Integrating these tracers into the reconstruction process would be interesting to examine.

In addition to using the reconstructed fields for parameter constraints, another promising application is their use in cross-correlation studies. For instance, inhomogeneous (patchy) reionization induces secondary fluctuations in the CMB. It is feasible to reconstruct the associated patchy $\tau$ map, which serves as a proxy for the ionization power spectrum integrated along the line of sight. According to Eq.~\eqref{eq:tb}, the 21-cm brightness temperature is sensitive to $x_{\rm HI}$. Cross-correlations between the brightness temperature and the reconstructed patchy $\tau$ signal have been explored in the literature \citep{Meerburg:2013dua,Roy:2019qsl}. However, the line-of-sight mode is typically lost when foregrounds are either avoided or removed, complicating the direct measurement of this cross-correlation. Based on our analysis, it is likely that the reconstructed $x_{\rm HI}$ field, even with large-scale cuts to account for foreground removal or avoidance, may be less affected by this issue. Moreover, the reconstructed $x_{\rm HI}$ field is expected to have a high coherence with the patchy $\tau$ signal, given that both measure the same underlying field, albeit one integrated along the line of sight. 

We are currently investigating this and other potentially valuable applications, which we aim to report in future publications.

\section*{Acknowledgements}
We link to thank Thomas Flöss, and Anchal Saxena for useful discussions and initial collaboration on the project. Furthermore, we like to thank the Center for Information Technology of the University of Groningen for their support and for providing access to the Habrok high-performance computing cluster. BB is supported by the Fundamentals of the Universe research program at the University of Groningen. LVEK acknowledges the financial support from the European Research Council (ERC) under the European Union’s Horizon 2020 research and innovation programme (Grant agreement No. 884760, "CoDEX").

\section*{Data Availability}
Accompanying code is available at \url{https://github.com/bidenkobd/reconstruction/}. The data underlying this article will be shared on reasonable request to the corresponding author.

\bibliographystyle{mnras}
\bibliography{bibtex}
\appendix

\section{cross-seed test}

\begin{figure*}
\centering 
 \includegraphics[width=1.\textwidth]{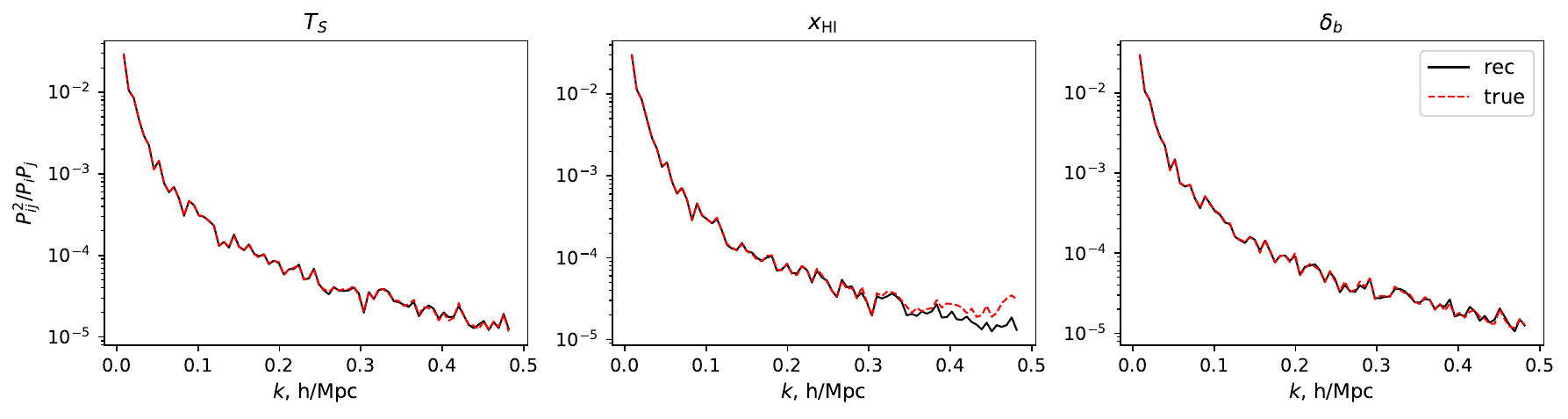}
 \caption{Coherence between simulated fields with different initial conditions (black) and corresponding reconstructed fields (red)}
 \label{fig:xseed}
\end{figure*}

When training our network, we use randomised seeds (phases) of the initial density field. This ensures that the network does not 'learn' a particular realisation of the fields but instead must learn the correlation structure. To verify that the randomisation of the seeds is sufficient and that the network does not pick up any recurring structure, we computed the correlation coefficients (coherence) of the true fields ($T_S$, $x_{\rm HI}$, and $\delta_b$) generated from different initial seeds and compared these to the correlation coefficients of the reconstructed fields, derived from applying the network to the simulated $\delta T_b$ sourced with randomised seeds. The results are shown in Fig.~\ref{fig:xseed}.

Overall, the correlation coefficient is $\ll 1$, as expected, given that these fields have different phases. Any correlation structure should be due to the chance correlation of the two random fields. The reconstructed fields also exhibit similar behaviour, where the chance correlations almost perfectly match those of the true fields. Only for the ionization field is there a small decrease in correlation after reconstruction on scales $k > 0.3 \, h \, \mathrm{Mpc}^{-1}$. 

The cause of this increase is not entirely clear. We have explored several hypotheses, but none adequately explains this behaviour. The effect is extremely small and may be simply a result of the coherence between the true field and the reconstructed fields dropping towards smaller scales. However, the absence of this effect in the other two reconstructed fields is somewhat puzzling. Further investigation is necessary to rule out any issues with the network fully. Because the effect is minor and has a negligible impact on the quality of the reconstruction, we will defer this investigation to future work.
construction, we will leave this for future work.  

\bsp
\label{lastpage}
\end{document}